%% file: main.tex
\documentclass[10pt,conference]{IEEEtran}
\IEEEoverridecommandlockouts
\usepackage{cite}
\usepackage{amsmath,amssymb,amsfonts}
\usepackage{algorithmic}
\usepackage{graphicx}
\usepackage{textcomp}
\usepackage{xcolor}
\usepackage{makecell}
\usepackage{multirow}
\usepackage{enumitem}
\usepackage{listings}

\definecolor{NGreen}{RGB}{0,176,80}
\definecolor{verylightgray}{rgb}{.97,.97,.97}
\lstdefinelanguage{Solidity}{
  keywords=[1]{anonymous, assembly, assert, balance, break, call, callcode, case, catch, class, constant, continue, constructor, contract, debugger, default, delegatecall, delete, do, else, emit, event, experimental, export, external, false, finally, for, function, gas, if, implements, import, in, indexed, instanceof, interface, internal, is, length, library, log0, log1, log2, log3, log4, memory, modifier, new, payable, pragma, private, protected, public, pure, push, require, return, returns, revert, selfdestruct, send, solidity, storage, struct, suicide, super, switch, then, this, throw, transfer, true, try, typeof, using, value, view, while, with, addmod, ecrecover, keccak256, mulmod, ripemd160, sha256, sha3}, 
  keywordstyle=[1]\color{blue}\bfseries,
  keywords=[2]{address, bool, byte, bytes, bytes1, bytes2, bytes3, bytes4, bytes5, bytes6, bytes7, bytes8, bytes9, bytes10, bytes11, bytes12, bytes13, bytes14, bytes15, bytes16, bytes17, bytes18, bytes19, bytes20, bytes21, bytes22, bytes23, bytes24, bytes25, bytes26, bytes27, bytes28, bytes29, bytes30, bytes31, bytes32, enum, int, int8, int16, int24, int32, int40, int48, int56, int64, int72, int80, int88, int96, int104, int112, int120, int128, int136, int144, int152, int160, int168, int176, int184, int192, int200, int208, int216, int224, int232, int240, int248, int256, mapping, string, uint, uint8, uint16, uint24, uint32, uint40, uint48, uint56, uint64, uint72, uint80, uint88, uint96, uint104, uint112, uint120, uint128, uint136, uint144, uint152, uint160, uint168, uint176, uint184, uint192, uint200, uint208, uint216, uint224, uint232, uint240, uint248, uint256, var, void, ether, finney, szabo, wei, days, hours, minutes, seconds, weeks, years},  
  keywordstyle=[2]\color{teal}\bfseries,
  keywords=[3]{block, blockhash, coinbase, difficulty, gaslimit, number, timestamp, msg, data, gas, sender, sig, value, now, tx, gasprice, origin},  
  keywordstyle=[3]\color{violet}\bfseries,
  identifierstyle=\color{black},
  sensitive=false,
  comment=[l]{//},
  morecomment=[s]{/*}{*/},
  commentstyle=\color{gray}\ttfamily,
  stringstyle=\color{red}\ttfamily,
  morestring=[b]',
  morestring=[b]"
}
\newcommand{\todo}[1]{}
\renewcommand{\todo}[1]{{\color{red} TODO: {#1}}}

\def\BibTeX{{\rm B\kern-.05em{\sc i\kern-.025em b}\kern-.08em
    T\kern-.1667em\lower.7ex\hbox{E}\kern-.125emX}}
\begin{document}

\title{Turn the Rudder: A Beacon of Reentrancy Detection for Smart Contracts on Ethereum}

\author{\IEEEauthorblockN{Zibin Zheng}
\IEEEauthorblockA{
  \textit{Sun Yat-sen University}, China \\zhzibin@mail.sysu.edu.cn}
\\
\IEEEauthorblockN{Zhijie Zhong}
\IEEEauthorblockA{
  \textit{Sun Yat-sen University}, China \\zhongzhj3@mail2.sysu.edu.cn}

\and
\IEEEauthorblockN{Neng Zhang*}
\IEEEauthorblockA{
  \thanks{*Neng Zhang is the corresponding author.}
  \textit{Sun Yat-sen University}, China \\zhangn279@mail.sysu.edu.cn}
\\
\IEEEauthorblockN{Mingxi Ye}
\IEEEauthorblockA{
  \textit{Sun Yat-sen University}, China \\yemx6@mail2.sysu.edu.cn}

\and
\IEEEauthorblockN{Jianzhong Su}
\IEEEauthorblockA{
  \textit{Sun Yat-sen University}, China \\sujzh3@mail2.sysu.edu.cn}
\\
\IEEEauthorblockN{Jiachi Chen}
\IEEEauthorblockA{
  \textit{Sun Yat-sen University}, China \\chenjch86@mail.sysu.edu.cn}
}

\maketitle

\begin{abstract}
  \input{section/abstract}
\end{abstract}


\section{Introduction}\label{intro}
\input{section/introduction.tex}

\section{Background}\label{background}
\input{section/preliminary}

\section{Methodology}\label{meth}
\input{section/methodology.tex}

\section{Results}\label{res}
\input{section/result.tex}

\section{Discussion}\label{dis}
\input{section/discussion.tex}

\section{Threats to Validity}\label{threats}
\input{section/threats.tex}

\section{Related Work}\label{relw}
\input{section/related_work.tex}

\section{Conclusions and Future Work}\label{con}
\input{section/conclusion.tex}

\section*{Acknowledgment}

This work is supported in part by the National Natural Science  Foundation of China (No. 62032025), the Key-Area Research and Development Program of Shandong Province (No. 2021CXGC010108), Special Projects in Key Fields of Universities in Guangdong Province (No. 2022ZDZX1001), and the Technology Program of Guangzhou, China (No. 202103050004)

\bibliographystyle{IEEEtran}
\bibliography{refs}

\end{document}

%% file: section/abstract.tex
Smart contracts are programs deployed on a blockchain and are immutable once deployed. Reentrancy, one of the most important vulnerabilities in smart contracts, has caused millions of dollars in financial loss. Many reentrancy detection approaches have been proposed. It is necessary to investigate the performance of these approaches to provide useful guidelines for their application. In this work, we conduct a large-scale empirical study on the capability of five well-known or recent reentrancy detection tools such as Mythril and Sailfish. We collect 230,548 verified smart contracts from Etherscan and use detection tools to analyze 139,424 contracts after deduplication, which results in 21,212 contracts with reentrancy issues. Then, we manually examine the defective functions located by the tools in the contracts. From the examination results, we obtain 34 true positive contracts with reentrancy and 21,178 false positive contracts without reentrancy. We also analyze the causes of the true and false positives. Finally, we evaluate the tools based on the two kinds of contracts. The results show that more than 99.8\% of the reentrant contracts detected by the tools are false positives with eight types of causes, and the tools can only detect the reentrancy issues caused by \emph{call.value()}, 58.8\% of which can be revealed by the Ethereum's official IDE, Remix. Furthermore, we collect real-world reentrancy attacks reported in the past two years and find that the tools fail to find any issues in the corresponding contracts. Based on the findings, existing works on reentrancy detection appear to have very limited capability, and researchers should turn the rudder to discover and detect new reentrancy patterns except those related to \emph{call.value()}.

%% file: section/introduction.tex
Smart contracts are programs deployed on a blockchain~\cite{zheng2020overview}. Due to the decentralized and trusted authorities guaranteed by blockchain technology, smart contracts are widely used to develop decentralized applications (DApps) in a variety of domains such as games, government, and finance~\cite{buterin2014next}. However, smart contracts are also restricted by the immutability of blockchain~\cite{hofmann2017immutability}. That is, a contract cannot be patched once deployed. A vulnerable contract could be leveraged by malicious attackers and result in serious problems, e.g., financial loss. Therefore, it is important to ensure the correctness of a contract before deploying it. This is a challenging task in practice for contract developers.

A number of vulnerabilities have been discovered for smart contracts from real-world attacks or through theoretical analysis~\cite{ji2020deposafe, ma2021pluto, ferreira2020aegis}. For example, 37 types of vulnerabilities are recorded in the SWC registry~\cite{SWC}, and the NCC Group~\cite{NCC} lists the top ten vulnerabilities, e.g., reentrancy and time manipulation. To enable developers to recognize and fix vulnerabilities, many approaches have been proposed to detect vulnerabilities in contracts~\cite{zhuang2020smart, bunz2020zether, ma2021pluto}. Reentrancy is a vulnerability that has been extensively studied by existing research as it could lead to huge financial loss, e.g., the DAO attack~\cite{mehar2019understanding} caused a loss of around 150 million dollars.

Existing reentrancy detection approaches mainly focus on the \emph{call.value()}\footnote{This is a typical ether transfer function in Solidity. The function may have different forms in different versions of Solidity. In this paper, we use call.value() for simplicity.} operations in smart contracts. In Section~\ref{background}, we explain a reentrancy issue using an example (see Fig.~\ref{fig:Reexp}). The reentrancy is caused by the delayed update of the state variable \emph{userbalance} behind the call to \emph{call.value()}. Reentrancy detection approaches aim to discover the payments that could be repeatedly incurred by external calls using various techniques such as symbolic execution~\cite{baldoni2018survey}, fuzzing~\cite{li2018fuzzing}, and neural networks~\cite{zhuang2020smart}. Two recent empirical studies~\cite{durieux2020empirical, xue2020cross} conducted two years ago revealed that there can be many false positives detected by the approaches at that time. The contracts used in these studies are written in Solidity version $\leq$0.6.0. In the past two years, the Solidity language has gone through several versions with significant changes~\cite{soliditydocs}. Inspired by these studies, a number of approaches have been upgraded or newly proposed. However, there has been no large-scale study on the performance of state-of-the-art reentrancy detection approaches on the contracts developed in Solidity version $>$0.6.0. Moreover, it has not yet been confirmed whether existing approaches could detect other reentrancy issues in spite of those related to \emph{call.value()}, as such reentrancy issues seem to be able to be detected by the official IDE, Remix~\cite{Remix}.

In this work, we conduct a large-scale empirical study on the capability of existing approaches in detecting reentrancy from smart contracts. We collect the verified Solidity code of all 230,548 contracts from Etherscan~\cite{Etherscan} (a leading block explorer and analytics platform for Ethereum) on October 13, 2021, and select five well-known or recent tools such as Mythril~\cite{mueller2018smashing} and Sailfish~\cite{bose2022sailfish} that can locate the possibly defective functions with reentrancy issues in contracts. After filtering duplicate contracts with the same bytecode, we obtain 139,424 contracts without duplication. Then, we use the tools to analyze the contracts, which results in 21,212 reentrant contracts. Next, we manually examine the defective functions of the reentrant contracts by recruiting 50 participants (including 27 undergraduates, 21 masters, and two PhDs). From the examination results, we build a set of 34 true positive contracts with reentrancy and a set of 21,178 false positive contracts without reentrancy. We also analyze the causes of the true and false positives. Using the two sets of contracts, we evaluate the tools. Furthermore, we test the tools on the contracts with reentrancy attacks reported in the past two years and also verify the true positive reentrant contracts using Remix. The results are as follows: 1) more than 99.8\% of the reentrant contracts detected by the tools are false positives with eight types of causes; 2) the true positive reentrant contracts are all related to \emph{call.value()}, 58.8\% of which can be discovered by Remix; and 3) the tools fail to detect reentrancy issues from the recently attacked contracts. Based on the results, we conclude that existing works on reentrancy detection have poor performance and may be outdated, and researchers should shift their attention from \emph{call.value()} to focus on discovering and detecting new reentrancy patterns. 

The main contributions of this work are outlined below:

\begin{itemize}
    \item We study the capability of five well-known or recent reentrancy detection tools on 139,424 smart contracts.
    \item We manually examine 21,212 reentrant contracts detected by the tools and build a set of 34 true positive contracts with reentrancy and a set of 21,178 false positive contracts without reentrancy. We further summarize eight types of causes that lead to false positives and find that all of the true positives are caused by \emph{call.value()}.
    \item We evaluate the tools based on the two manually built sets of true and false positive reentrant contracts. Moreover, we verify the true positives using Remix and test the tools on the contracts with recent reentrancy attacks. Based on the results, we provide insightful guidelines for researchers.
    \item We release our experimental data at GitHub~\cite{EData}, including the 230,548 contracts, the detection results of the tools, and the two sets of true and false positive contracts, to provide a benchmark for researchers to conduct future work on reentrancy detection.
\end{itemize}

The rest of the paper is organized as follows. Section~\ref{background} introduces smart contracts and reentrancy vulnerability. Section~\ref{meth} describes our research methodology. Section~\ref{res} presents the results. Section~\ref{dis} discusses the results and describes two additional tests. Section~\ref{threats} analyzes the threats to validity of our study. Section~\ref{relw} reviews the related work. Section~\ref{con} concludes the paper and discusses future work.

%% file: section/preliminary.tex
\begin{figure}
	\centering
	\begin{lstlisting}[language=Solidity,mathescape, firstnumber=1,escapechar=\%]
contract SimpleDAO {
    mapping (address => uint) public userbalance;
    ...
    function withdraw(uint amount) public{
        if (userbalance[msg.sender]>= amount) {
            require(msg.sender.call.value(amount)());
            userbalance[msg.sender]-=amount;
        }
    }  
}
	\end{lstlisting}
	\caption{Simple example of reentrancy}
	\label{fig:Reexp}
\end{figure}

\subsection{Smart Contracts}
Smart contracts are programs running on a blockchain. In general, smart contracts are written in Solidity~\cite{soliditydocs}, one of the most popular languages for smart contracts. The Solidity code of a smart contract is compiled into bytecode and then deployed on blockchain. In addition, the compilation process generates the application binary interface (ABI) to facilitate subsequent calls and analysis of smart contracts.

\subsection{Reentrancy}
From the 150 million USD DAO attack in 2016 to the 80 million USD Fei Protocol attack in 2022, the reentrancy vulnerability has caused huge financial loss. In this section, we describe how a reentrancy attack could happen with a simplified version of the DAO smart contract, as shown in Fig.~\ref{fig:Reexp}. The example contract is developed for asset management. It uses the variable \textit{userbalance} (line 2) to record the balance of each user and allows users to call the \textit{withdraw()} function (line 4) to withdraw their balance. In \textit{withdraw} function, the contract first checks if the caller (represented by the address variable \textit{msg.sender}) has enough balance (line 5); then, it transfers the requested \textit{amount} of ether to the caller and subtracts the \emph{amount} from the caller's balance recorded in the variable \textit{userbalance}. However, Solidity introduces a special mechanism named the ``fallback function''. Users can write their own code in the fallback function and the function will be executed if the contract receives ether from other addresses. In the example case, the ether transfer function \textit{call.value()} (line 6) will automatically call the fallback function of the caller contract and thus the caller can take over the control flow. Attackers can deploy malicious code in the fallback function to repetitively invoke the \textit{withdraw()} function. Note that in the second invocation of \textit{withdraw()}, line 7 has not been executed since the invocation begins at the \textit{call.value()} function in line 6, and thus the \emph{userbalance} has not been changed at this time. As a consequence, the condition check (line 5) of the second invocation is passed and the victim contract will transfer ether to the caller repeatedly until the balance of the contract is drained.

%% file: section/methodology.tex
\begin{figure*}
    \centering
    \includegraphics[width=0.85\linewidth]{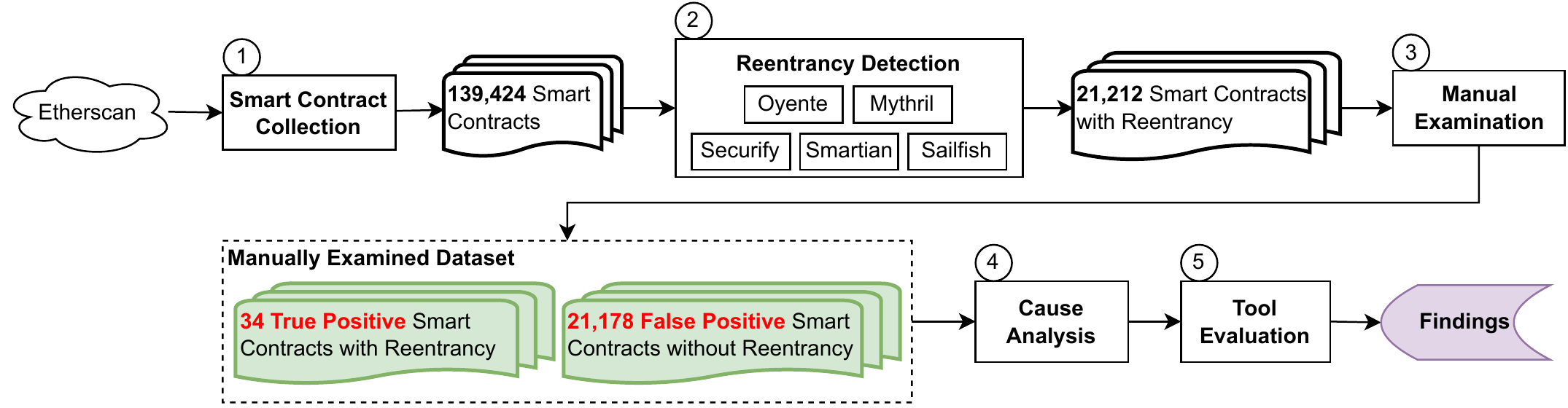}
    \caption{The overview of our research methodology}
    \label{fig:met}
\end{figure*}

As shown in Fig.~\ref{fig:met}, our research methodology contains five main steps: 1) \emph{Smart Contract Collection}, which collects smart contracts from Etherscan; 2) \emph{Reentrancy Detection}, which detects reentrancy issues from the contracts using five selected tools; 3) \emph{Manual Examination}, which manually examines whether the detected reentrant contracts are correct (i.e., true positive) or not (i.e., false positive); 4) \emph{Cause Analysis}, which analyzes the causes of the true and false positive contracts; and 5) \emph{Tool Evaluation}, which evaluates the tools based on the manually examined and analyzed results from the contracts.

\subsection{Smart Contract Collection}\label{sc_col}

We collect 230,548 smart contracts with verified Solidity code, ABI, and bytecode from Etherscan on October 13, 2021. Through preliminary analysis, there are a considerable number of duplicate contracts with the same bytecode. To reduce the workload of the subsequent steps of our study, we filter out the duplicates. Consequently, we obtain 139,424 contracts without duplication. Table~\ref{tab: collected-dataset} summarizes the information of the contracts, including their average lines of code (LOC) and the distribution of their versions of Solidity compilers.

\begin{table}[t]
    \centering
        \caption{Summaries of our smart contract dataset.}
        \resizebox{0.85\linewidth}{!}{
        \begin{tabular}{l|c}
            \hline
            \textbf{Information} & \textbf{Number} \\
            \hline
            \#Contracts & 230,548\\
            \#Deduplicated contracts & 139,424 \\
            \hline
            \#Contracts with compiler version \textless 0.5 & 68,196 \\
            \#Contracts with compiler version 0.5+ & 23,461 \\
            \#Contracts with compiler version 0.6+ & 24,761 \\
            \#Contracts with compiler version 0.7+ & 8,292 \\
            \#Contracts with compiler version 0.8+ & 14,714 \\
            \hline
            Average LOC in Deduplicated Contracts & 720.42 \\
            \hline
        \end{tabular}}
        \label{tab: collected-dataset}
\end{table}
    
\subsection{Reentrancy Detection}

\subsubsection{Tool Selection}\label{tool_sel}
To conduct our study, we need to select some representative tools that can detect reentrancy issues from contracts. In the empirical study conducted by Durieux et al.~\cite{durieux2020empirical}, they summarize a list of 35 vulnerability detection tools for smart contracts. We extend their list by searching for other state-of-the-art tools published in the literature or on the internet after that work, e.g., Smartian~\cite{choi2021smartian}, Sailfish~\cite{bose2022sailfish}, etc. However, not all of the tools are appropriate for our study, e.g., those that cannot detect reentrancy or cannot be applied to Solidity code. To select appropriate tools for this study, we define several criteria as follows. 

\begin{itemize}
    \item \textbf{Available and scalable}. The tool is publicly available and can be easily applied to a large set of contracts. In particular, the tool should support a command-line interface that is convenient for performing large-scale experiments.
    \item \textbf{Supports multiple versions of Solidity}. The tool can detect the contracts in our dataset that are written in multiple versions of Solidity.
    \item \textbf{Requires Solidity code only}. The tool only requires the Solidity code and its derivatives (e.g., ABI and Bytecode) as input without any other specification (e.g., a test suite annotated with assertions).
    \item \textbf{Ability to locate vulnerabilities}. The tool can locate the defective functions with reentrancy issues in a contract, which is important for practical use and can facilitate the manual examination task in this study.
\end{itemize}

Using the criteria above, we select five representative reentrancy detection tools from both industry and academia, as listed in Table~\ref{tab: tool-docker}. Within industry, we select Mythril since it has demonstrated its performance in~\cite{durieux2020empirical}. Within academia, we select Oyente, Securify, Smartian, and Sailfish from the top conferences of software engineering or security. The selected tools use various techniques, e.g., symbolic execution, formal verification, and fuzzing, which are briefly described below:
    
\textbf{Oyente}~\cite{luu2016making} is one of the first smart contract analysis tools based on symbolic execution. It constructs the control flow graph of a contract and symbolically executes the contract to detect vulnerabilities by exploring as many execution paths as possible. Oyente serves as the basis of several other vulnerability detection tools such as Maian~\cite{nikolic2018finding} and Osiris~\cite{torres2018osiris}.
    
\textbf{Mythril}~\cite{mueller2018smashing} is also a vulnerability detection tool based on symbolic execution, which combines taint analysis and control flow checking for more accurate detection. Particularly, Mythril has been packaged as a commercial product by Consensys~\cite{Consensys}.

\textbf{Securify}~\cite{tsankov2018securify} is a vulnerability detection tool based on formal verification. It symbolically analyzes the dependency graph of a contract and checks compliance/violation patterns that capture sufficient conditions for proving whether a property holds or not. Note that there are two versions of Securify, namely Securify(V1) and Securify(V2). We use both of them because they support different versions of Solidity.
    
\textbf{Smartian}~\cite{choi2021smartian} is a fuzzer that combines static analysis and dynamic analysis to detect vulnerabilities from contracts. It statically analyzes a contract to predict the transaction sequences that can lead to effective testing, and then uses such information to construct the initial seed corpus. During fuzzing, Smartian performs a lightweight dynamic data-flow analysis to effectively guide fuzzing.

\textbf{Sailfish}~\cite{bose2022sailfish} is a scalable system for automatically finding state inconsistency bugs in smart contracts. In order to make the analysis tractable, Sailfish contains two phases: a lightweight exploration phase for reducing the number of instructions to analyze, and a refinement phase for generating extra constraints to approximate the side effects of whole-program execution and ensure the precision of the symbolic evaluation. Using these phases, Sailfish can efficiently detect state inconsistencies in smart contracts.
    
\subsection{Experiment Setup}
We use the selected tools to analyze smart contracts using Docker images, as listed in Table~\ref{tab: tool-docker}. We use two methods to obtain the Docker images. For Oyente, Mythril, Securify(V1), and Sailfish, we directly download their images from Dockerhub~\cite{Dockerhub}. For Securify(V2) and Smartian, there are no images on Dockerhub, and so we build up their images according to the Dockerfiles\footnote{A Dockerfile is a text document that contains all the commands a user could call on the command line to assemble an image.} in their GitHub repositories. We notice that Securify(V1) has multiple Docker images on Dockerhub, and different images can support the analysis of contracts with different versions of Solidity. We use multiple Docker images to increase the analysis range of the contracts.

We directly use Oyente, Mythril, Securify and Sailfish on the Solidity code of each contract. Smartian takes the bytecode and ABI of a contract as input, which are also collected from Etherscan. Referring to the time budget set in~\cite{durieux2020empirical}, we set an appropriate time budget, i.e., two minutes, for the five selected tools per contract. If the time budget is up, the tool stops the analysis and exports the analysis result.

Notice that some of the five selected tools do not report the detected reentrancy issues as reentrancy. For example, Mythril reports two kinds of vulnerabilities related to reentrancy, namely \emph{External Call To User-Supplied Address} and \emph{State access after external call}. We determine the reentrancy related vulnerabilities reported by the tools based on several kinds of original materials, including the papers, homepages, implementation code, and output of the tools, which are provided in our repository~\cite{EData}. For instance, the ``DAO'' reported by Securify is selected as a reentrancy related vulnerability as the comments in the implementation code indicate that ``DAO'' represents reentrancy. Table~\ref{tab: tool-docker} lists the vulnerabilities related to reentrancy.

\begin{table*}[t]
    \centering
        \caption{Five representative reentrancy detection tools used in our study}
        \resizebox{0.9\linewidth}{!}{
            \begin{tabular}{l|l|l|l|l}
                \hline
                \textbf{Tool} & \textbf{Technique} & \textbf{GitHub Repository URL} & \textbf{Docker} & \textbf{Reentrancy Patterns} \\ \hline
                Oyente & \begin{tabular}[c]{@{}l@{}}Symbolic \\ execution\end{tabular} & https://github.com/enzymefinance/oyente & qspprotocol/oyente-0.4.25 & Re-Entrancy Vulnerability \\ 
                \hline
                Mythril & \begin{tabular}[c]{@{}l@{}}Symbolic \\ execution\end{tabular} & https://github.com/ConsenSys/mythri & mythril/myth & \begin{tabular}[c]{@{}l@{}}External Call To User-Supplied Address\\ State access after external call\end{tabular} \\ 
                \hline
                \multirow{2}{*}{Securify} & \multirow{2}{*}{\begin{tabular}[c]{@{}l@{}}Formal\\ verification\end{tabular}} & V1: https://github.com/eth-sri/securify & \begin{tabular}[c]{@{}l@{}}qspprotocol/securify-0.4.25\\ qspprotocol/securify-usolc\end{tabular} & \begin{tabular}[c]{@{}l@{}}DAO \\ DAOConstantGas\end{tabular} \\ \cline{3-5} & & V2: https://github.com/eth-sri/securify2 & Dockerfile & \begin{tabular}[c]{@{}l@{}}Benign Reentrancy\\ Reentrancy with constant gas\\ Gas-dependent Reentrancy\\ No-Ether-Involved Reentrancy\end{tabular} \\ 
                \hline
                Sailfish & \begin{tabular}[c]{@{}l@{}}Formal \\ verification\end{tabular} & https://github.com/ucsb-seclab/sailfish & holmessherlock/sailfish:latest & DAO \\ 
                \hline
                Smartian & Fuzzing & https://github.com/SoftSec-KAIST/Smartian & Dockerfile & Reentrancy \\ 
                \hline
                \end{tabular}
        }
        \label{tab: tool-docker}
\end{table*}

\subsection{Manual Examination}\label{subsec:man_exam}
In order to perform a deep analysis of the tools used in this study, we need to examine whether the detected reentrant contracts are correct. Recall that our selected tools can locate the defective functions in a contract, which can facilitate the examination task. In total, there are 31,720 defective functions detected from 21,212 contracts. Manually examining such a large number of functions is a heavy workload.

\subsubsection{Participant Recruitment}\label{par_rec}
To reduce the workload of the manual examination task, we need to recruit a relatively large number of participants and let each participant examine only a subset of contracts. Given a defective function of a contract, in order to judge whether the function has a reentrancy issue, the participants should be equipped with knowledge about Solidity and the reentrancy mechanism. It is not an easy job to recruit sufficient participants and ensure that they meet the requirements. In the lead co-author's affiliation, there are more than one hundred undergraduates and masters who could be potential participants. However, not all of the students are familiar with Solidity and reentrancy. Fortunately, in the first co-author's research group, there are a number of PhD students concentrated on research on the vulnerability detection of smart contracts, with 2-5 years of experience. The PhDs have good knowledge of Solidity and reentrancy. With the help of two experienced PhDs, we adopt a four-stage process to recruit participants from the undergraduates and masters as follows.

\begin{itemize}
    \item \textbf{Invitation.} We send an invitation email to 70 undergraduates and 25 masters. In the email, we introduce our study and the manual examination task, and also ask the students whether they are willing to participate in the task. One week later, we receive positive feedback from 36 undergraduates and 23 masters.
    \item \textbf{Training}. We launch several online conferences with the 59 students and invite two PhDs with experience in smart contracts to introduce Solidity and the reentrancy mechanism with examples. Both PhDs also introduce two true positive reentrancy patterns and five false positive reentrancy patterns that they have either discovered from prior studies (e.g.,~\cite{xue2020cross}) or by themselves.
    \item \textbf{Self-learning.} After the training stage, we ask the students to further learn Solidity and reentrancy for two weeks by themselves using some suggested materials (e.g., the official Solidity documentation and online tutorials on reentrancy) or web search. We encourage the students to develop, deploy, and execute a contract with reentrancy using Remix.
    \item \textbf{Testing.} After the self-learning stage, we conduct a test to see whether the students have gained enough knowledge of Solidity and reentrancy to perform the examination task. We collect 15 contracts, of which five contracts have reentrancy issues. We use our selected tools on the contracts. The possibly defective functions with reentrancy in the five contracts are detected by at least one of the tools. We develop a website to facilitate the examination task for the students. The website randomly presents each contract in a set with the corresponding list of defective functions detected from the contract. We ask the students to examine the defective functions of a contract one by one and annotate a function that has a reentrancy issue with a specific string ``$\langle$yes$\rangle$ Reentrancy''. After the test is completed, we evaluate the accuracy of each student's annotations. A total of 27 undergraduates and 21 masters achieve an accuracy of 100\%, and they are chosen as participants for the manual examination task.
\end{itemize}

\subsubsection{First Round of Manual Examination}\label{subsec:man_exam1}
We randomly divide the 48 participants into 24 groups. Each group contains two participants. We also randomly divide the 21,212 reentrant contracts detected by the tools into 24 subsets. 23 subsets contain 884 contracts, and one subset contains 880 contracts. Each subset is allocated to a participant group. The participants do not know which group they belong to. The aforementioned website is used to facilitate the manual examination task. The website randomly displays each contract in the subset allocated to a participant, along with a list of defective functions detected from the contract. When the participants click a defective function, the website can quickly skip to the function in the contract. We ask the participants to examine the defective functions one-by-one and annotate a function that has a reentrancy issue with a specific string ``$\langle$yes$\rangle$ Reentrancy''.

Once the manual examination task is completed, we obtain three sets of contracts: 1) $commPs$: the set of contracts that have at least one function annotated by both participants in a group; 2) $commNs$: the set of contracts that have no function annotated by both participants in a group; and 3) $Diffs$: the set of contracts with different annotations given by both participants in a group (i.e., one participant annotates at least one function while the other participant does not annotate any function). There are 97, 20,626, and 489 contracts contained in $commPs$, $commNs$, and $Diffs$, respectively.

\subsubsection{Review of the Manual Examination Results}
Considering that the participants may not be proficient in smart contracts, we ask the two experienced PhDs involved in the training process to review the participants' examination results. We randomly sample 377 contracts from $commNs$, which is a statistically significant sample size considering a confidence level of 95\% and a confidence interval of 5\%. Since there is no annotation added to the contracts in $commNs$, both PhDs need to examine the defective functions of the contracts by themselves. They independently examine each of the contracts. After the examination process, both PhDs do not annotate any function of the contracts, meaning that the participants' judgements on the contracts are all correct. This result probably thanks to the introduction of false positive patterns of reentrancy during the training process. Next, we ask both PhDs to review the participants' annotations of the 97 contracts in $commPs$. They perform the review independently. If the annotations of a contract are correct, they label the contract as 1; otherwise 0. Consequently, there are 83 contracts labeled as 0 by both PhDs, indicating that the participants fail to accurately identify reentrancy issues from the contracts. Based on the two groups of results, we are confident in the annotations of contracts in $commNs$, while we lack confidence in the annotations of contracts in $commPs$ and the contracts with disagreement in $Diffs$. 

Although the reentrancy issues identified by the participants are not reliable, the participants accurately identify a large number of false positive contracts without reentrancy, which greatly reduces the number of contracts that need to be subsequently examined by the experienced PhDs.

\begin{table*}[t]
	\centering
	\caption{The analysis results of five vulnerability detection tools. }
	\resizebox{\linewidth}{!}{
		\renewcommand{\arraystretch}{1.3}
		\begin{tabular}{cc|cc|cc|cc|cc|cc|cc|cc}
			\hline
			& & \multicolumn{2}{c}{\textbf{Oyente}}  & \multicolumn{2}{c}{\textbf{Mythril}} & \multicolumn{2}{c}{\textbf{Securify (V1)}} & \multicolumn{2}{c}{\textbf{Securify (V2)} } & \multicolumn{2}{c}{\textbf{Smartian} } & \multicolumn{2}{c}{\textbf{Sailfish} } & \multicolumn{2}{c}{\textbf{Total} } \\
			\textbf{Version} & \textbf{Num.} & \textbf{RE} & \textbf{Analyzed} & \textbf{RE} & \textbf{Analyzed} & \textbf{RE} & \textbf{Analyzed} & \textbf{RE} & \textbf{Analyzed} & \textbf{RE} & \textbf{Analyzed} & \textbf{RE} & \textbf{Analyzed} & \textbf{RE} & \textbf{Analyzed} \\
			\hline
            \textless 0.5 & 68,196 & 513 & 56,289 & 11,962 & 40,006 & 2,324 & 48,715 & 0 & 37 & 15 & 59,511 & 1,403 & 43,541 & 14,967 & 66,170 \\ 
            0.5+ & 23,461 & 0 & 576 & 1,877 & 15,954 & 63 & 13,761 & 1,693 & 15,726 & 3 & 18,075 & 641 & 17,401 & 3,805 & 22,350 \\ 
            0.6+ & 24,761 & 0 & 54 & 1,263 & 14,988 & 0 & 127 & 797 & 4,497 & 2 & 15,356 & 244 & 4,789 & 2,040 & 20,470 \\ 
            0.7+ & 8,292 & 0 & 7 & 273 & 3,363 & 0 & 7 & 2 & 71 & 2 & 4,155 & 1 & 61 & 275 & 5,542 \\ 
            0.8+ & 14,714 & 0 & 13 & 125 & 7,035 & 0 & 29 & 0 & 44 & 0 & 3,148 & 0 & 14 & 125 & 8,683 \\ 
			\hline
			\textbf{Total} & 139,424 & 513 & 56,939 & 15,500 & 81,346 & 2,387 & 62,639 & 2,492 & 20,375 & 22 & 100,245 & 2,289 & 65,806 & 21,212 & 123,215 \\
			\hline
		\end{tabular}
	}
	\label{tab: output-of-tools}
\end{table*}

\subsubsection{Second Round of Manual Examination}
According to the review results above, we ask the two experienced PhDs to re-examine the 586 (97+489) contracts in $commPs$ and $Diffs$ using a card sorting approach~\cite{CardSorting}. Both PhDs first independently examine and annotate the defective functions of each contract, similar to the examination task described in Section~\ref{subsec:man_exam1}. After the examination process, there are 15 contracts with different annotations. By discussing the disagreements together, both PhDs reach a consensus. Finally, we obtain a set of 34 true positive contracts with reentrancy, denoted as $TPs$, and a set of 21,178 (20,626+586-34) false positive contracts without reentrancy, denoted as $FPs$.

\subsection{Cause Analysis}\label{cause_ana}
To better understand the true and false positive contracts, we further ask the two PhDs to analyze the causes of the contracts. To reduce the workload, we randomly sample 377 contracts from $FPs$, which is a statistically significant sample size considering a confidence level of 95\% and a confidence interval of 5\%. However, it is difficult to include any of the 15 false positives reported by Smartian in the sample as those false positives only occupy a tiny part of the entire set of 21,178 false positives. Therefore, we directly include the 15 false positives reported by Smartian in the scope of cause analysis. In total, we analyze the causes of 392 (377+15) false positives. Since we do not have a predefined set of all possible causes, we ask both PhDs to perform cause analysis using two substeps. One PhD first analyzes the cause of the 426 (392+34) contracts and records the cause using a short description. The other PhD then reviews the recorded causes. In cases of disagreement, both PhDs discuss the cause to reach a common decision. From the analysis results, the 34 true positive contracts are all related to the \emph{call.value()} function, while there are eight types of causes that lead to the 392 false positive contracts (see Table~\ref{tab:falsePositiveTools}).

\subsection{Tool Evaluation}
Using the two sets of true and false positive contracts with different causes, we evaluate the five tools used in this study. In spite of the overall performance in terms of precision (see Table~\ref{tab:precision}), we perform a deep analysis by counting the numbers of true and false positive contracts detected by the tools, with respect to each cause type, as listed in Table~\ref{tab:falsePositiveTools}.

%% file: section/result.tex
In this section, we present and discuss the results of the analytical procedure described in Section~\ref{meth}.

\subsection{Automated Analysis Results} \label{sec: result-of-tool}
Table~\ref{tab: output-of-tools} presents the automated analysis results of the five tools on our dataset. For each tool, the first column (RE) is the number of reentrant contracts reported by the tool, and the second column (Analyzed) is the number of contracts successfully analyzed by the tool. 

\textbf{Analysis Scope.} Solidity is a new and frequently updated language; the new features, along with the frequent updates, limit the scope of analysis of vulnerability detection tools. In particular, the versions of Solidity above 0.5.0 have major changes in their grammar~\cite{soliditydocs}, so that the analysis scope of some tools (e.g., Oyente, Securify(V1), and Securify(V2)) is limited in specific versions. Accordingly, we divide our dataset (139,424 smart contracts) into five parts, as shown in Table~\ref{tab: output-of-tools}.

Oyente and Securify(V1) fail to analyze most of the smart contracts above version 0.5.0, and Securify(V2) can only analyze smart contracts above version 0.5.8. Sailfish fails to analyze smart contracts with a version above 0.7.0. Note that there are a few smart contracts above version 0.5.0 that can be successfully analyzed by Oyente, which might be because the contracts do not use the new features of the later versions. This phenomenon also appears with Sailfish, Securify(V1) and Securify(V2). In particular, the analysis scopes of Mythril and Smartian are less affected by the version of a smart contract, being able to successfully analyze most of the smart contracts in our dataset. In addition to the limitation of versions, the other reason for analysis to fail is that the analysis does not successfully terminate when an external timeout is reached, in which case we kill the process. Consequently, there are 15,778, 1,891, and 2,522 contracts that have no results exported by the tools Mythril, Securify(V1), and Securify(V2). The successful analysis rates of the tested tools show a decreasing trend when applied to higher versions of Solidity, which is ranging from 97\% ($\frac{66,170}{68,196}$) in \textless 0.5 to 60\% ($\frac{8,683}{14,714}$) in 0.8+.

In order to analyze as many contracts as possible, we combine the automated analysis results of all tools for manual checking. However, there still exist 16,209 smart contracts that fail to be analyzed by any tool.

\textbf{Reported Reentrancy Rate.} As listed in Table~\ref{tab: output-of-tools}, the five tools report a total of 21,212 reentrant contracts. The reported rate, i.e., $\frac{\#reported\_reentrant\_contracts}{\#successfully\_analyzed\_contracts}$, varies by tool. Mythril successfully analyzes 81,346 contracts and 19\% are reported as reentrancy issues. These 15,500 issues occupy 78\% of all reported reentrancy issues. Smartian successfully analyzes the greatest number of smart contracts but reports the smallest number of reentrancy issues. There are only 22 reported reentrancy issues in 100,245 contracts that are successfully detected by Smartian. This is likely because the patterns used to detect reentrancy are different. These patterns are decisive for vulnerability detection tools. In the next section, we summarize and propose some effective patterns for vulnerability detection tools to detect reentrancy issues.

\textbf{Reported Reentrancy Distribution.} From Table~\ref{tab: output-of-tools}, we find that the reported reentrancy rate decreases as the version of Solidity updates, ranging from 22.6\% ($\frac{14,967}{66,170}$) in \textless 0.5 to 1.4\% ($\frac{125}{8,683}$) in 0.8+. There may be three reasons for this: 1) the developers' awareness of preventing reentrancy issues is strengthened, as many companies have been victims of the reentrancy vulnerability and several related works have been proposed for detecting it; 2) codebases for preventing reentrancy have been proposed and are widely used, such as the \textit{ReentrancyGuard} proposed by openzepplin~\cite{reentrancyguard}; and 3) the new types of reentrancy are hard for existing tools to detect. In the next section, we conclude and discuss some new types of reentrancy that may enable developers to write safer contracts or propose more effective vulnerability detection tools.

\subsection{Precision of Tools}

\begin{table}[t]
	\centering
	\caption{The number of reentrant contracts reported (Reported Num.) by five tools, and the number of true positives (TP Num.).}
	\begin{tabular}{c|c|c}
		\hline
		\textbf{Tool} & \textbf{TP Num.} & \textbf{Reported Num.} \\
		\hline
		\textbf{Oyente} & 25 & 513 \\
		\textbf{Mythril} & 26 & 15,500 \\
		\textbf{Securify(V1)} & 15 & 2,387 \\
		\textbf{Securify(V2)} & 3 & 2,492 \\
		\textbf{Smartian} & 7 & 22 \\
		\textbf{Sailfish} & 19 & 2,289 \\
		\hline
		\textbf{Total} & 34 & 21,212 \\
		\hline
	\end{tabular}
	\label{tab:precision}
\end{table}

According to the manual examination, we evaluate the precision of the detection tools. As shown in Table~\ref{tab:precision}, all tools have very low precision in revealing reentrancy vulnerabilities. Smartian achieves the highest precision, i.e., 31.8\%, among these tools. However, Smartian only reports 22 contracts with reentrancy, of which seven contracts are true positives. In general, the precision of the tools is poor for real-world applications in detecting reentrancy vulnerabilities.

\subsection{False Positives of Tools}
\label{subsec: FPofTools}
In this subsection, we summarize the common reasons for the false positives of the tested tools. 

\subsubsection{Permission control}
Symbolic execution-based tools, e.g., Oyente and Mythril, usually use a specific pattern to detect vulnerabilities. However, they fail to consider the user's permission when checking a control flow path. There are three ways that could lead to permission control-based false positives, i.e., identify control, address control, and reentrancy lock (a defense mechanism against reentrancy).

\noindent {\bf 1. Identity Control:} The first type of false positive is caused by the ignorance with regard to identity control of the contract caller. Fig.~\ref{fig:ACmodifier} shows a code snippet of a real-world smart contract that is falsely detected to have reentrancy vulnerability by Oyente and Mythril. The ether transfer function \emph{call.value()} in line 8 of the contract is detected to lead to reentrancy vulnerability. The function execution enables the caller to transfer the amount of \textit{\_value} ether to the address \textit{\_to}. Malicious users could attack this smart contract from the fallback function in the smart contract deployed at the \textit{\_to} address if they can call the example function. However, this function is not callable from arbitrary addresses due to the \textit{onlyOwner} modifier at line 1, which allows only the contract owner to execute the function. Therefore, the function can not be reentered by other malicious smart contracts.

\begin{figure}
	\centering
	\begin{lstlisting}[language=Solidity,mathescape, firstnumber=1]
	modifier onlyOwner{
	    require(msg.sender == owner);
	    _;
	}
	...
	function execute( address _to, uint _value, bytes _data) external onlyOwner {
	    ...
	    _to.call.value(_value)(data);
	}
	\end{lstlisting}
	\caption{Code example: Identify control based permission control}
	\label{fig:ACmodifier}
\end{figure}

\noindent {\bf 2. Address Control:}
The code snippet in Fig.~\ref{fig:ACaddress} shows a case of false positive type, which is caused by the limited access control of the contract address. This contract is detected to have reentrancy vulnerability in the \textit{register()} function. There is an external call in line 6, where the function calls the \textit{transferFrom()} function from the contract in the address \textit{dai}. However, when we look into the address variable, it can be seen that the address \textit{dai} is assigned with a hard code address in line 2 and line 3, which cannot be modified by other functions. Therefore, the smart contract corresponding to the address \textit{dai} is the smart contract recorded in the address on Ethereum. As a result, the external call in line 6 can not be reentered by other malicious smart contracts.

\begin{figure}
	\centering
	\begin{lstlisting}[language=Solidity,mathescape, firstnumber=1]
	contract DaiSavingsEscrow{
	    address private daiAddress = 0x6B175474E89094C44Da98b954EedeAC495271d0F;
	    IERC20 public dai = IERC20(daiAddress);
    	function register(...) public {
        	...
        	dai.transferFrom(msg.sender, vault, payment);
        	...
        }
    }
	\end{lstlisting}
	\caption{Code example: Address based permission control}
	\label{fig:ACaddress}
\end{figure}

\noindent {\bf 3. Reentrancy Lock:}
Fig.~\ref{fig:AClock} demonstrates another case of a false positive caused by reentrancy lock. The function in line 15 is detected to have a reentrancy vulnerability based on the external call in line 17. It seems that a smart contract corresponding to the \textit{Token} address could deploy malicious code to reenter the \textit{withdraw()} function to obtain additional profit. However, the function is actually safe against reentrancy attacks due to the modifier \textit{nonReentrant}. This modifier is declared in line 6, where the modifier checks the state of a variable \textit{\_notEntered} before the execution of the function and sets it to be ``false'' (line 7 and 8). If the check fails, the \textit{require} statement will stop the execution of this transaction and roll back to the state before the transaction executes. The ``\_'' in line 9 is a placeholder for the body of the function and line 10 will only be executed after the execution of the function body is finished. Note that before the execution of line 10, the value of variable \textit{\_notEntered} is ``false'' during the execution of the function body. As a result, if any external smart contract tries to reenter \textit{withdraw()} for the second time before finishing the first execution, the check statement in line 7 will fail and the transaction will be stopped immediately. Therefore, the smart contract is safe against reentrancy attacks.

\begin{figure}
	\begin{lstlisting}[language=Solidity,mathescape, firstnumber=1]
contract ReentrancyGuard {
    bool private _notEntered;
    constructor () internal {
        _notEntered = true;
    }
	modifier nonReentrant() {
    	require(_notEntered);
    	_notEntered = false;
    	_;
    	_notEntered = true;
    }
}
contract GovernanceVesting is ReentrancyGuard {
    ...
	function withdraw() public nonReentrant {
    	...
    	IERC20(Token).transfer(governanceAddress, governanceFunds);
    	Withdrawn = true;
    }
}
	\end{lstlisting}
	\caption{Code example: Reentrancy lock based permission control}
	\label{fig:AClock}
\end{figure}

\subsubsection{No State Change After External Call}
If a function could be executed by external calls several times in a single transaction, the contract is regarded as reentrant contract by the previous reentrancy detection tools. However, in some situations, the external calls may not involve financial acts and no state will be changed after the external call. For example, Fig.~\ref{fig:NoStateChange} declares a function \textit{getTokenBal()} to inquire the balance of address \textit{who} in the token contract \textit{tokenAddr} (line 9). However, the state of the contract (i.e., ether balance, storage variable, etc.) will not be changed after the external call of the \textit{balanceOf()} function. Therefore, even if a malicious token \emph{t} reenters this function, it cannot affect the execution of the smart contract.

\begin{figure}
	\begin{lstlisting}[language=Solidity,mathescape, firstnumber=1]
	contract ForeignToken {
    	function balanceOf(address _owner) constant public returns (uint256);
    	...
    }
	contract Bitcash {
    	...
    	function getTokenBal(address tokenAddr, address who) constant public returns (uint){
        	ForeignToken t = ForeignToken(tokenAddr);
        	uint bal = t.balanceOf(who);
        	return bal;
        }
    }
	\end{lstlisting}
	\caption{Code example: No state change after external call}
	\label{fig:NoStateChange}
\end{figure}

\subsubsection{Change State Variable without Financial Risk}
Some smart contracts may change state variables after the external call, making them potentially vulnerable to reentrancy attacks. However, not all state changes lead to reentrancy vulnerability. Take the function in Fig.~\ref{fig:NoFinancialRisk} as an example. The code in line 3 defines an external call to address `\textit{token}', and then the code in line 4 assigns it to a state variable \textit{tokens}. There is no following code in the function, and this function is not called in any other functions in the contract\footnote{The full Solidity code of the contract can be found in https://etherscan.io/address/0x046ec9bb312c51425f7a00b2ab7525afe7db52e6}. Reentering this function in line 3 will not cause financial risk because the \textit{transferFrom()} function is to transfer tokens from the first parameter (\textit{msg.sender} in this case) to the second parameter (\textit{this} smart contract in the case). Thus, \textit{transferFrom()} is for the function caller to transfer tokens to this smart contract, and reentering this function only increases the balance of this smart contract. In this case, no financial risk exists in the contract.

\begin{figure}
	\begin{lstlisting}[language=Solidity,mathescape, firstnumber=1]
	function depositToken(address token, uint amount) {
    	...
    	if (!Token(token).transferFrom(msg.sender, this, amount)) throw;
    	tokens[token][msg.sender] = safeAdd(tokens[token][msg.sender], amount);
    }
	\end{lstlisting}
	\caption{Code example: Change state variable without financial risk}
	\label{fig:NoFinancialRisk}
\end{figure}

\begin{table*}[t]
	\centering
	\caption{Cause analysis results of the false positives detected by five tools. The last column `Total' is the union of the false positives reported by the tools.}
	\begin{tabular}{l|cccccc|c}
		\hline
		\textbf{Cause Type} & \textbf{Oyente} & \textbf{Mythril} & \textbf{Securify(V1)} & \textbf{Securify(V2)} & \textbf{Smartian} & \textbf{Sailfish} & \textbf{Total} \\
		\hline
		Permission Control (Identity Control)                             & 6   & 152  & 12   & 12    & 8     & 1  & 178  \\
		Permission Control (Address Control)                    & 2   & 16  & 2     & 14    & 2     & 3  & 32\\
		Permission Control (Reentrancy Lock)                       & 0   & 2   & 0     & 6     & 0     & 0  &6\\
		No State Change After External Call             & 2  & 99   & 2     & 14    & 1     & 0  & 117  \\
		Change State Variable without Financial Risk    & 6  & 12   & 8     & 21    & 1     & 3  & 44   \\
		Special Transfer Value                          & 0  & 4    & 0     & 7     & 2     & 1  & 12   \\
		Reentrancy by Transfer/Send                     & 0  & 1    & 0     & 8     & 1     & 0  & 10  \\
		Non-callable Function                           & 2  & 7    & 25    & 1     & 0     & 1  & 33  \\
		\hline
	\end{tabular}
	\label{tab:falsePositiveTools}
\end{table*}

\subsubsection{Special Transfer Value}
This type of false positive is caused by the ignorance of parameter semantics in the ether transfer function. For example, the function in Fig.~\ref{fig:SpecialValue} is detected to be reentrant due to the ether transfer function in line 5. However, the amount parameter is \textit{msg.value}, which means the ether transferred to the external address \textit{wethToken} is the ether amount received by the function \textit{tradeEthVsDAI}. In this sense, reentering this function would not cause financial loss since it does not affect the balance of the smart contract.

\begin{figure}
	\begin{lstlisting}[language=Solidity,mathescape, firstnumber=1]
	function tradeEthVsDAI(uint numTakeOrders, uint numTraverseOrders, bool isEthToDai, uint srcAmount) public payable {
    	...
    	if (isEthToDai) {
        	require(msg.value == srcAmount);
        	wethToken.deposit.value(msg.value)();
        	...
    	} else ...
    }
	\end{lstlisting}
	\caption{Code Example: Special Transfer Value}
	\label{fig:SpecialValue}
\end{figure}

\subsubsection{Reentrancy by transfer/send}
The gas system is a special mechanism introduced by Ethereum to limit the resource consumption of smart contracts. Once the smart contract runs out of gas, the execution will be terminated and all states will be rolled back. Unlike the \emph{call.value()} function, \textit{transfer()} and \textit{send()} will change the maximum gas limitation to 2,300 units when the recipient is a contract. For the \textit{transfer()} function in Fig. \ref{fig:ReenTransfer}, the 2,300 gas limits is insufficient for a call to contract or a write operation to any storage variable, which means that the attackers cannot raise reentrancy attack.

\begin{figure}
	\begin{lstlisting}[language=Solidity,mathescape, firstnumber=1]
	function _withdraw(address from,address payable to,address token,uint256 amount) internal {    
	    ...
    	if(token == address(0)) {  
    	    to.transfer(amount);
    	} else ...
    }  
	\end{lstlisting}
	\caption{Code example: Reentrancy by transfer/send}
	\label{fig:ReenTransfer}
\end{figure}

\subsubsection{Non-callable Function}
There are two types of bytecode in Ethereum: runtime bytecode and creation bytecode. Creation bytecode contains the information that will never be executed after the contract is deployed on a blockchain, e.g., the constructor function. However, Mythril fails to identify this situation. When the source code is used as the input, Mythril simply compiles the source code to creation bytecode and uses it for vulnerability detection. This gap between creation bytecode and runtime bytecode results in a type of false positive located in the constructor function. Since this type of function will not be recorded on the blockchain, it will never be called and be attacked by malicious attackers. Besides, as shown in Fig.~\ref{fig:ReenTransfer}, the "internal" functions are not available to be called by other contracts, which are also reported as reentrant contracts by some tools.

\subsection{Distribution of False Positives in Each Tool}
According to the sampled false positives, we present the distribution of the false positives reported by each tool in Table~\ref{tab:falsePositiveTools}. We have the following observations: 1) The false positives related to \textit{Permission Control} occupy the most, i.e., 55\% ($\frac{178+32+6}{392}$). This is because there are several ways to deal with the permission of the contracts, which are hard for analyzers to handle. 2) The tools report two types of false positives, \textit{No State Change After External Call} and \textit{Change State Variable without Financial Risk}, which occupy 41\% ($\frac{117+44}{392}$). These two types relate to the functions in other contracts, which increases the difficulty of analysis. As for other types of false positives (i.e., \textit{Special Transfer Value}, \textit{Reentrancy by Transfer/Send} and \textit{Non-callable Function}), they occupy 14\% ($\frac{12+10+33}{392}$) of the false positives.

%% file: section/discussion.tex
\subsection{Reentrancy in the wild}

\begin{table}[t]
	\centering
	\caption{Real Reentrancy attacks on Ethereum}
	\resizebox{1.0\linewidth}{!}{
	\begin{tabular}{p{10pt}<{\centering}|p{25pt} <{\centering}|p{20pt}<{\centering}|p{35pt}<{\centering}|p{120pt}}
		\hline
		\textbf{No.} & \textbf{Time} & \textbf{Lost} & \textbf{Attacked Projects} & \textbf{Description} \\
		\hline
		1 & 2020/04 & \$3.5M& Uniswap &  A vulnerability when adopting ERC777 tokens\\
		\hline
		2 & 2020/04 & \$25M& Lendf.Me & A vulnerability when adopting ERC777 tokens \\
		\hline
		3 & 2020/11 & \$2M& Akropolis & A combination of reentrancy attack and flash loan attack to exploit the saving pools  \\
		\hline
		4 & 2020/11 & \$8M& OUSD & A combination of reentrancy attack and flash loan attack by utilizing mint logic flaw in its contracts\\
		\hline
		5 & 2021/07 & \$0.1M& DeFiPie & A combination of reentrancy attack and flash loan attack to withdraw almost all available liquidity from the protocol \\
		\hline
		6 & 2021/08 & \$18M& Cream Finance & A combination of reentrancy attack and flash loan attack to exploit a vulnerability in AMP token contract\\
		\hline
		7 & 2022/03 & \$1M& Bacon Protocol & A logic error in its lend() routine to allow hacker to get more lending credits\\
		\hline
		8 & 2022/03 & \$0.1M& Revest Finance & Bad design in the minting-related functions — do not strictly follow the check-validation-interaction model when transferring the ERC1155 token.
		
		  \\
		\hline
	\end{tabular}}
	\label{fig:real_reentrancy}
\end{table}

With the development of the smart contract ecosystem, reentrancy vulnerability has become much rarer in recent years. First, dozens of tools are developed to detect reentrancy. For example, the Ethereum's official IDE (Remix) warns of risk when the check-effects-interaction pattern is detected in a smart contract. In addition, as one of the most famous attacks on Ethereum, the DAO attack (the first smart contract reentrancy attack) has been widely introduced in smart contract tutorials, e.g., books, blogs, and videos. Thus, developers are well educated in avoiding reentrancy caused by \textit{call.value().}

Unfortunately, most academic works still focus on reentrancy caused by \textit{call.value()}, which might be the wrong direction based on today's Ethereum ecosystem. Based on the SlowMist hacked repository\footnote{https://hacked.slowmist.io/en/}, only eight real reentrancy attacks happened in Ethereum from Apr, 2020 to Apr, 2022, as shown in Table~\ref{fig:real_reentrancy}. It is not difficult to find that these real reentrancy attacks are much more complicated compared with the traditional \textit{call.value()} related reentrancy issues detected by most tools. These eight attacks could be classified into three groups. First, the No. 1, 2, and 8 reentrancy attacks are caused by bad designs when using ERC tokens~\cite{norvill2019standardising}, e.g., ERC777 and ERC1155. The bad design provides opportunities for hackers to transfer the tokens in these contracts. Second, the No. 3-6 reentrancy attacks should be combined with a flash loan\footnote{Flash loans allow users to borrow and settle loans instantaneously in a single transaction without providing any collateral.}, as this kind of reentrancy attack usually needs a large amount of capital to affect some key values of smart contracts. For example, buying almost all the tokens in a liquidity pool might lead to a logic error in contracts. Third, the No. 7 reentrancy attack was led by a logic error in the contract. In this case, attackers could obtain more lending credits than they paid.

The aforementioned reentrancy attacks are very complicated. Detecting them needs a high level of professional skill and a deep understanding of reentrancy attacks, rather than using a simple detection pattern. Evidence can be found in the case of the Akropolis smart contract (No. 3 attack), which was audited by two professional smart contract audit teams. However, neither of them found the reentrancy issues. This shows that some real reentrancy attacks are much more complicated and need more advanced methods to be detected. 

\subsection{IDE Warning of Reentrancy}
With the development of the smart contract language, the Ethereum's official IDE, i.e., Remix, has an integrated static analyzer to check for vulnerabilities, bad development practices, etc. The analyzer provides warnings for potential reentrancy vulnerability if the check-effects-interaction pattern is detected in the smart contract. A natural question is whether the detected true positive contracts could be warned when developing them with Remix. We test all of the 34 true positives using Remix, and find that 20 (i.e., 58.8\%)  of them are warned to have a potential reentrancy vulnerability.

%% file: section/threats.tex
\textbf{Threats to internal validity} relate to the errors in the implementation of the reentrancy detection tools and the bias of participants in the manual examination of the reentrant contracts detected by the tools and the cause analysis of true (\emph{resp.} false) positive contracts with (\emph{resp.} without) reentrancy. We implement the tools using their Docker images or Dockerfiles released at GitHub. For the manual examination task, we perform two rounds to reduce the heavy workload of 21,212 contracts. In the first round, we recruit 48 participants willing to participate in our task and ensure that they have relatively sufficient knowledge of Solidity and reentrancy to perform the task, adopting a four-stage process (Section~\ref{par_rec}). In the second round, we ask two PhDs with experience in Solidity and reentrancy to review the examination results of the first round and re-examine the contracts with low-quality results. For the cause analysis task, we ask both PhDs to collaboratively analyze the causes of true and false positive contracts. To avoid the subjective bias of a single person, the manual examination and cause analysis tasks for each contract are performed by two participants (including the two PhDs). It should be pointed out that the participants might miss some possible reentrancy issues in contracts that they do not know or have not been trained with.

\textbf{Threats to external validity} relate to the generalizability of the results. The objective of this study is to investigate the capability of existing works on reentrancy detection for smart contracts. We collect a large dataset of 230,548 contracts with Solidity code from Etherscan and select five well-known or recent reentrancy detection tools according to some key criteria for practical use (Section~\ref{tool_sel}). We exclude some popular tools as they either cannot be applied to Solidity code or cannot locate the possibly defective functions in contracts. We acknowledge that our results obtained using five tools cannot reflect all existing works on reentrancy detection. However, the results can help researchers gain a good understanding of the limitations of existing works and motivate researchers to address new reentrancy issues in spite of those related to \emph{call.value()}.

%% file: section/related_work.tex
\subsection{Reentrancy Detection Tools}
Reentrancy is one of the most notorious vulnerabilities, first discovered in 2016~\cite{luu2016making}, and continually discussed since. Due to the significance of this security issue, many detection tools have been proposed to prevent smart contracts from being attacked, including dynamic testing tools and static analyzers. For example, ContractFuzzer~\cite{jiang2018contractfuzzer}, sFuzz~\cite{nguyen2020sfuzz}, and Smartian~\cite{choi2021smartian} use dynamic testing~\cite{hanson1993testing} technologies to trigger reentrancy issues. ContractFuzzer is a fuzzing tool, which develops test oracles~\cite{baresi2001test} for the reentrancy vulnerability. Due to its inefficiency, Nguyen et al. propose an adaptive strategy in sFuzz to guide fuzzer toward executing unreached paths. However, these tools pay little attention to the characteristics of smart contracts. Choi et al. claim that some paths can only be reached by critical transaction sequences and thus propose Smartian. 

Oyente~\cite{luu2016making}, Securify~\cite{tsankov2018securify}, Slither~\cite{feist2019slither}, and Zeus~\cite{kalra2018zeus} use static analysis technologies to discover reentrancy vulnerability. In 2016, Luu et al. propose Oyente which addresses the problem of reentrancy detection. Oyente constructs the control flow graph of contracts and identifies reentrancy issues by risky patterns~\cite{luu2016making}. Due to the lack of context information, Oyente is imprecise. Tsankov et al. propose Securify to extract semantic information from smart contracts and perform analysis to check the existence of predefined patterns. Furthermore, Feist et al. propose Slither to perform automatic analysis~\cite{feist2019slither}. Bose et al. put forward Slither to handle state-inconsistency bugs (e.g., reentrancy) and achieve a better performance~\cite{kalra2018zeus}.

\subsection{Benchmark in Smart Contracts}
A benchmark is a foundation for comparing different methods~\cite{perazzi2016benchmark}. However, only a few benchmarks have been proposed in the area of smart contracts. \textit{VeriSmartBench}~\cite{verismartbench} is one of these benchmarks, which contains examples for all CVE cases. Unfortunately, this repository contains very few examples of reentrancy issues. \textit{SolidiFi-benchmark}~\cite{durieux2020empirical} is another benchmark which contains 31 examples with reentrancy issues, but many of them are artificial and toy examples.

Efforts have been made in other empirical studies to find different aspects of insight. Durieux et al.~\cite{durieux2020empirical} conduct a large-scale evaluation of automated analysis tools. Rather than focusing on specific vulnerabilities, they pay close attention to the effectiveness and efficiency of distinct tools. In addition, Perez et al.~\cite{perez2021smart} focus on evaluating whether vulnerable contracts have been exploited. They reveal that only a small proportion of smart contract vulnerabilities have been profited from. Xue et al.~\cite{xue2020cross} reveal five \textit{path protective techniques} which are relevant with the false alarm of reentrancy detection. Yet, these approaches only cover a portion of Solidity versions and cannot be used to draw a comprehensive conclusion.

%% file: section/conclusion.tex
In this paper, we conduct a large-scale empirical study to investigate the performance of existing works on reentrancy detection for smart contracts. We collect 230,548 contracts from Etherscan and select five well-known or recent tools that can locate the possibly defective functions with reentrancy issues in a contract. By manually examining 21,212 reentrant contracts detected using these tools, we obtain 34 true positive contracts with reentrancy and 21,178 false positive contracts without reentrancy. We also analyze the causes of the true and false positives. Using the two sets of contracts with different causes, we evaluate the tools. Two additional tests are conducted to evaluate the tools on a number of contracts with recent reentrancy attacks and to verify whether the detected reentrancy issues can be identified by the Ethereum's official IDE, Remix. The results show that the tools have a extremely high false positive rate of more than 99.8\% and that the tools can only detect reentrancy related to \emph{call.value()}, 58.8\% of which can be detected by Remix. Based on the results, we suggest that researchers turn to discovering and detecting new reentrancy issues in spite of the classical and simple patterns related to \emph{call.value()}. In future work, we plan to improve existing reentrancy detection tools by reducing false positives based on our summarized causes, e.g., adding rule-based filters to the tools by automating some of the patterns described in Section~\ref{subsec: FPofTools}. Moreover, we will study the reentrancy issues reported in recent attacked contracts.